 \documentclass[preprint2]{aastex}

\shorttitle{Be abundances in cool main-sequence stars with exoplanets}

\shortauthors{Delgado Mena et al.}

\begin{document}




\title{Be abundances in cool main-sequence stars with exoplanets\thanks{Based on
observations made with UVES at VLT Kueyen 8.2 m telescope at the European
Southern Observatory (Cerro Paranal, Chile) in program 86.D-0082A}}







\author{E. Delgado Mena\altaffilmark{1,2}, G. Israelian\altaffilmark{1,2}, J. I.
Gonz\'alez Hern\'andez\altaffilmark{1,2}, N. C. Santos\altaffilmark{3,4} and R.
Rebolo\altaffilmark{1,2,5}}

\altaffiltext{1}{Instituto de Astrofisica de Canarias, 38200 La Laguna,
Tenerife, Spain: edm@iac.es}

\altaffiltext{2}{Departamento de Astrof\'isica, Universidad de La Laguna, 38206
La Laguna, Tenerife, Spain}

\altaffiltext{3}{Centro de Astrof\'isica, Universidade do Porto, Rua das
Estrelas, 4150-762 Porto, Portugal}

\altaffiltext{4}{Departamento de F\'isica e Astronomia, Faculdade de Ci\^encias,
Universidade do Porto, Portugal}

\altaffiltext{5}{Consejo Superior de Investigaciones Cient\'{\i}ficas, Spain}






\begin{abstract}

We present new UVES spectra of a sample of 15 cool unevolved stars with and
without detected planetary companions. Together with previous
determinations, we study Be depletion and possible differences in Be abundances
between both groups of stars. We obtain a final sample of 89 and 40 stars with
and without planets, respectively, which covers a wide range of effective
temperatures, from 4700 K to 6400 K, and includes several cool dwarf stars for
the first time.

We determine Be abundances for these stars and find that for most of them (the
coolest ones) the BeII resonance lines are often undetectable, implying
significant Be depletion. While for hot stars Be abundances are aproximately
constant, with a slight fall as T$_{\rm eff}$ decreases and the Li-Be gap around
6300 K, we find a steep drop of Be content as T$_{\rm eff}$ decreases for
T$_{\rm eff}$ $<$ 5500 K, confirming the results of previous papers. Therefore,
for these stars there is an unknown mechanism destroying Be that is not
reflected in current models of Be depletion.

Moreover, this strong Be depletion in cool objects takes place for all the stars
regardless of the presence of planets, thus, the effect of extra Li depletion in
solar-type stars with planets when compared with stars without detected planets
does not seem to be present for Be, although the number of stars at those
temperatures is still small to reach a final conclusion.

\end{abstract}






\section{Introduction}

Light elements are important tracers of stellar internal mixing and rotation.
Since they are burned at relatively low temperatures they constrain 
how the material inside stars is mixed with the hotter
interior. Rotation and angular momentum loss are among the leading processes to
explain the mixing that leads to depletion of light elements in
solar-type stars \citep[e.g.][]{stephens,Bouvier} although gravitational waves
may also affect the abundances of those elements \citep{Montalban}. However, 
available models for evolution of Be (considering rotation) do not predict a 
significant depletion of Be during the main sequence for stars with 6000 K $>$ 
T$_{\rm eff}$ $>$ 4000 K \citep{Pinsonneault}. On the other hand, models which take 
into account gravitational waves predict significant Be depletion only for stars 
cooler than 4500 K \citep{Montalban}.\\

In a recent work, \citet{israelian09} confirmed that Li was severely depleted in
solar-type stars (with T$_{\rm eff}$ between 5600 K and 5850 K) with planets
when compared with similar stars without detected planets although this result
is controversial \citep{Baumann,sousa10}; for a complete discussion see
\citet{Delgado11}. This difference in Li abundance seems to be related to the
different rotational history of both groups of stars due to the presence of
protoplanetary disks \citep[e.g.][]{Bouvier}. However, beryllium needs
a greater temperature to be burned so we would expect to see the onset
of this effect in cooler stars where convective envelopes are deep enough to
reach those higher temperatures.\\

In a previous paper, \citet{Delgado11} found two cool planet host stars
with an extra depletion of Be when compared with analog stars without detected
planets. This encourages us to try to investigate this process in cool stars. In
this work we continue that analysis by extending the sample with 15 new cool
stars. We refer the reader to that paper for further information and a more
extensive introduction.\\

\begin{figure*}[ht!]
\centering
\includegraphics[width=16cm]{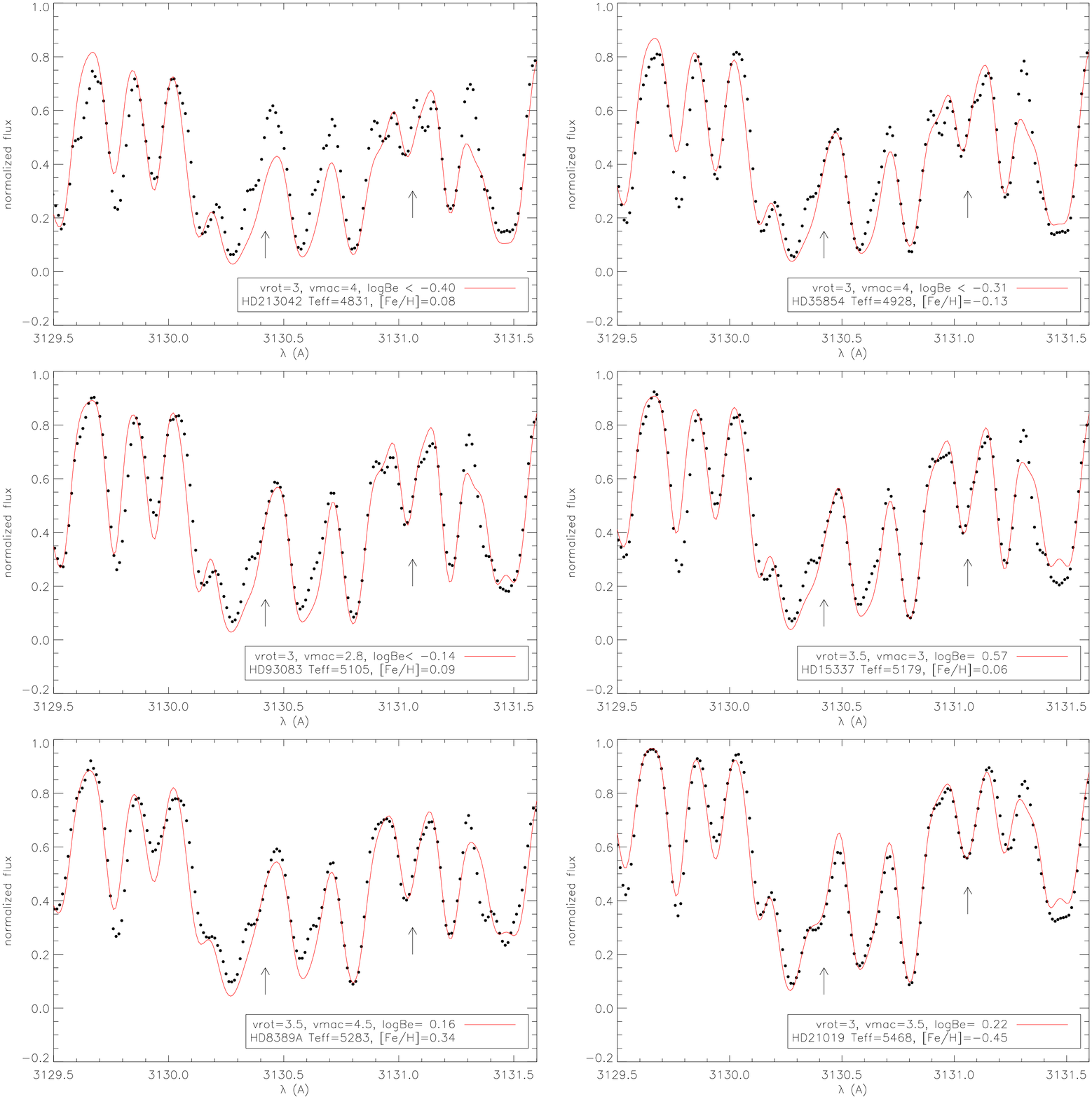}
\caption{Synthetic spectra (red lines) and observed spectra (dots) for the
planet host star HD 93083 and the stars without detected planets HD 213042, HD
35854, HD 15337, HD 8389A and HD 21019. The position of Be lines are indicated
by the arrows.}
\label{ajustes_be}
\end{figure*}











\begin{figure*}[ht!]
\centering
\includegraphics[width=8cm]{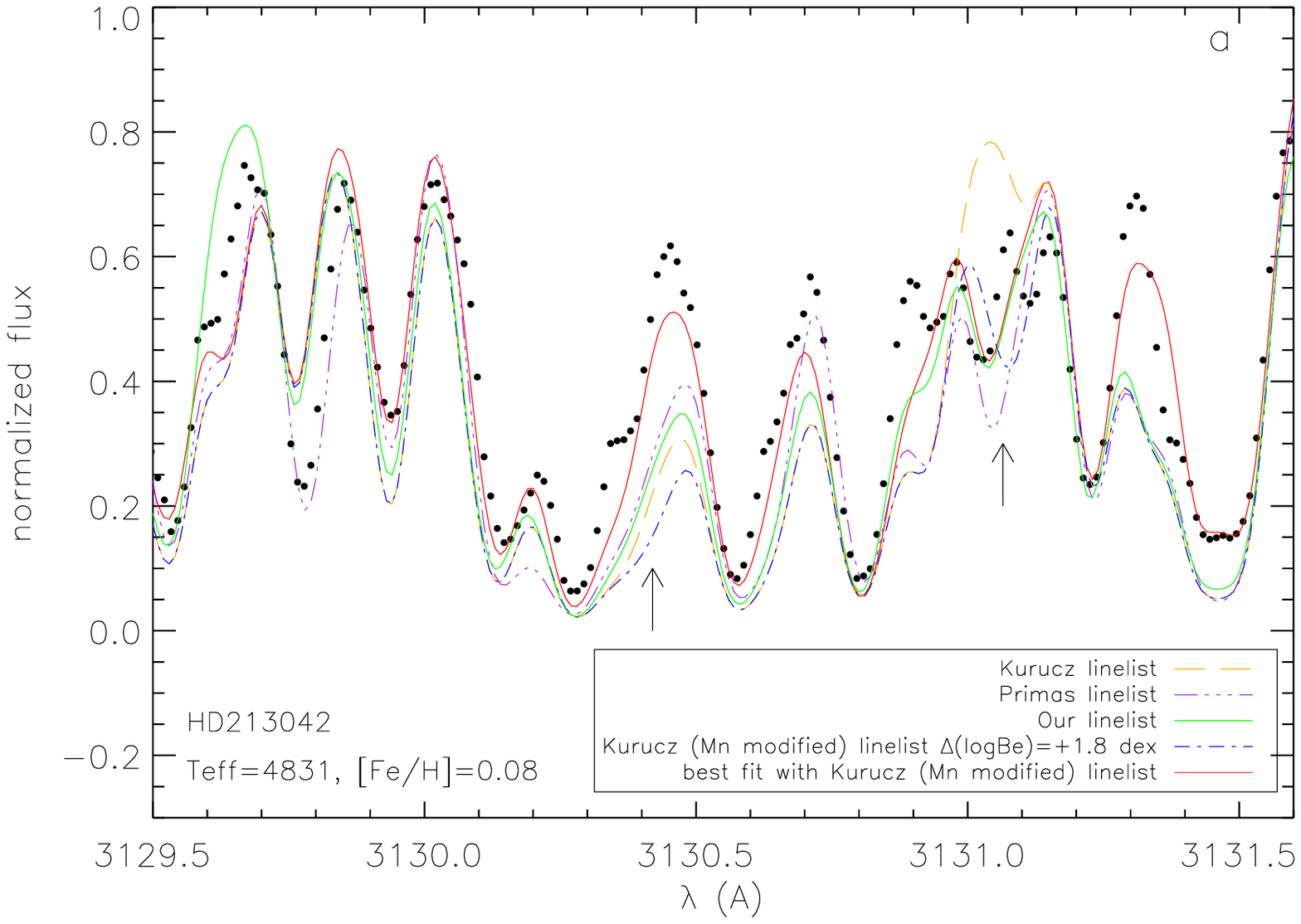}
\includegraphics[width=8cm]{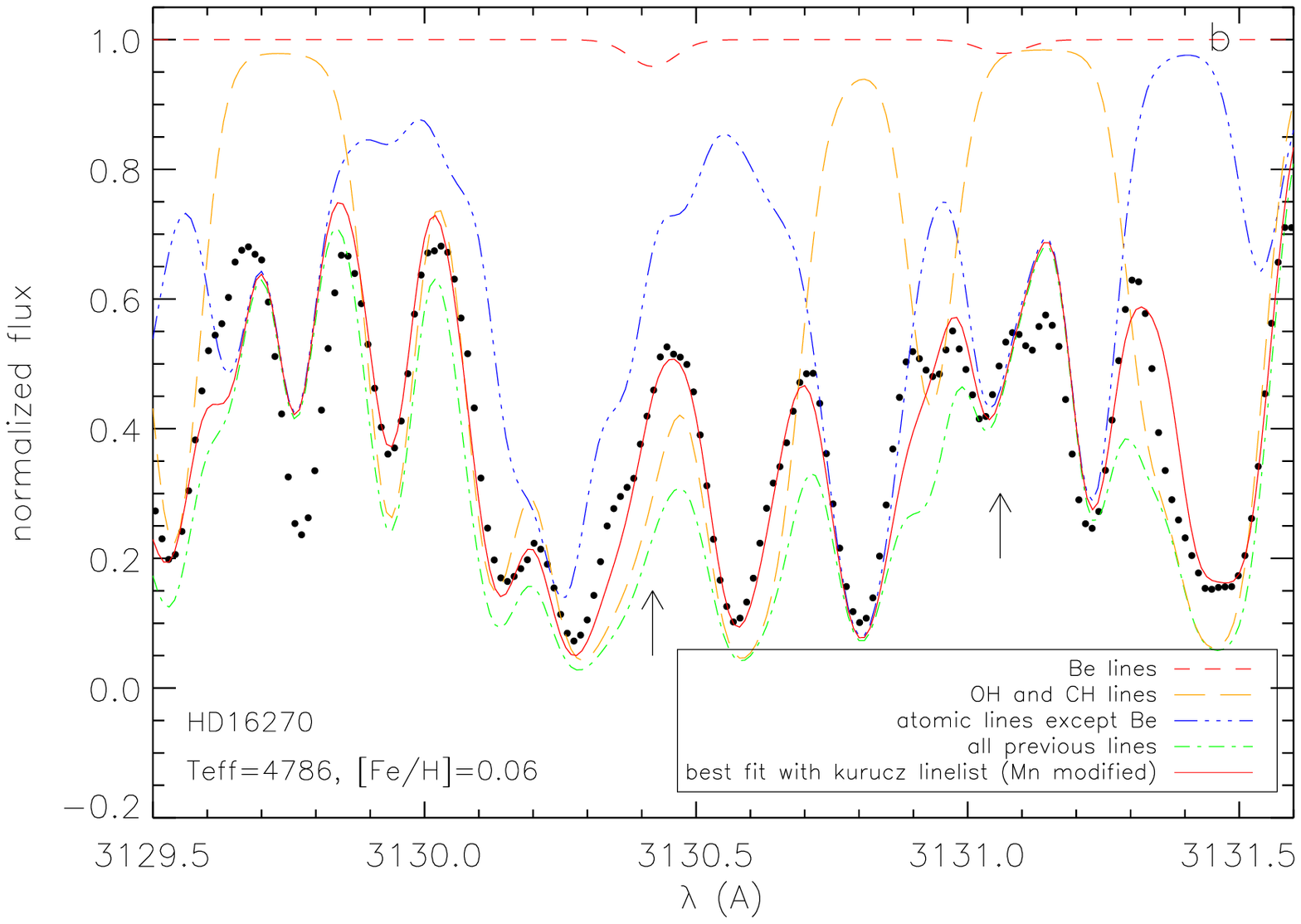}
\includegraphics[width=8cm]{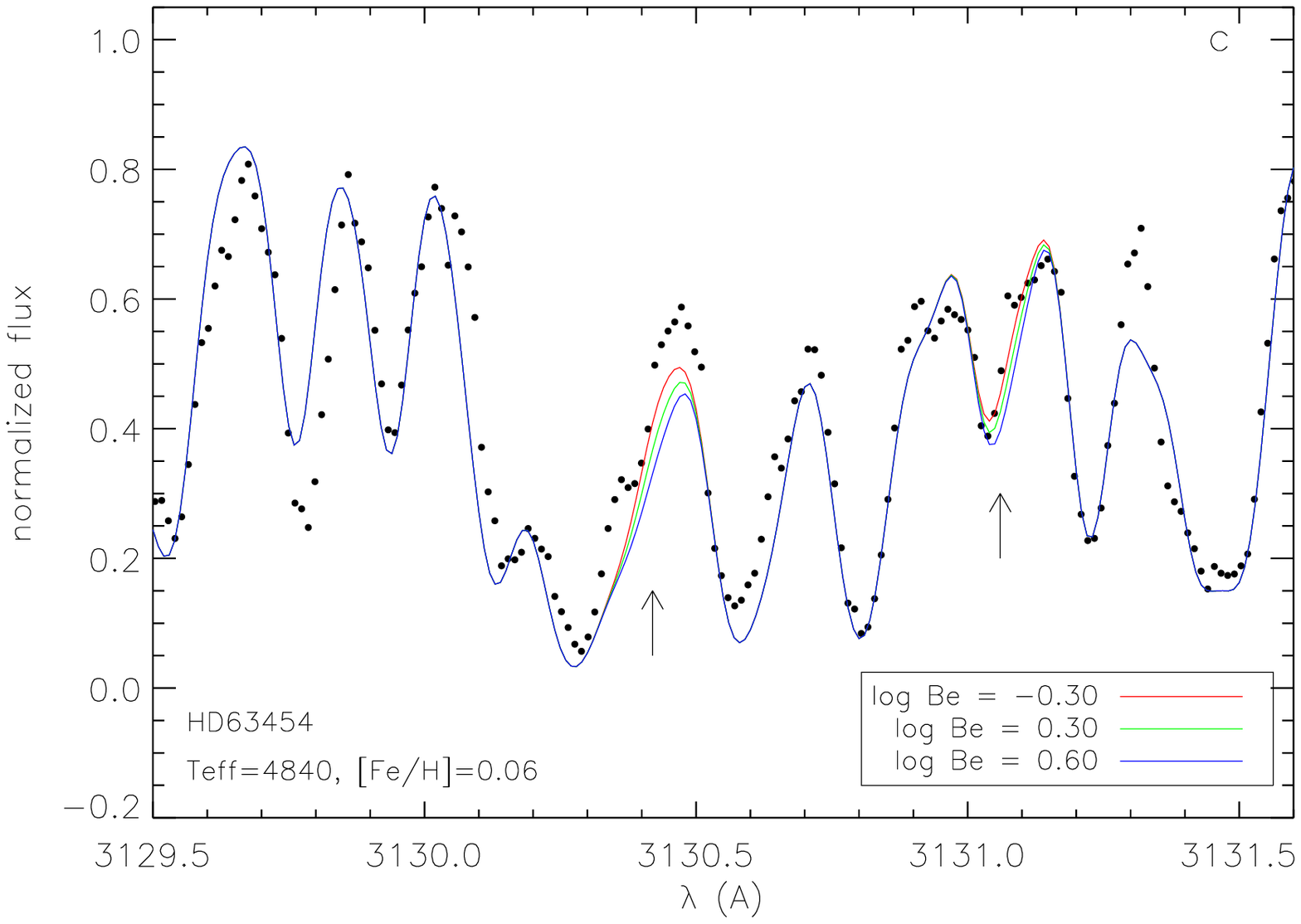}
\includegraphics[width=8cm]{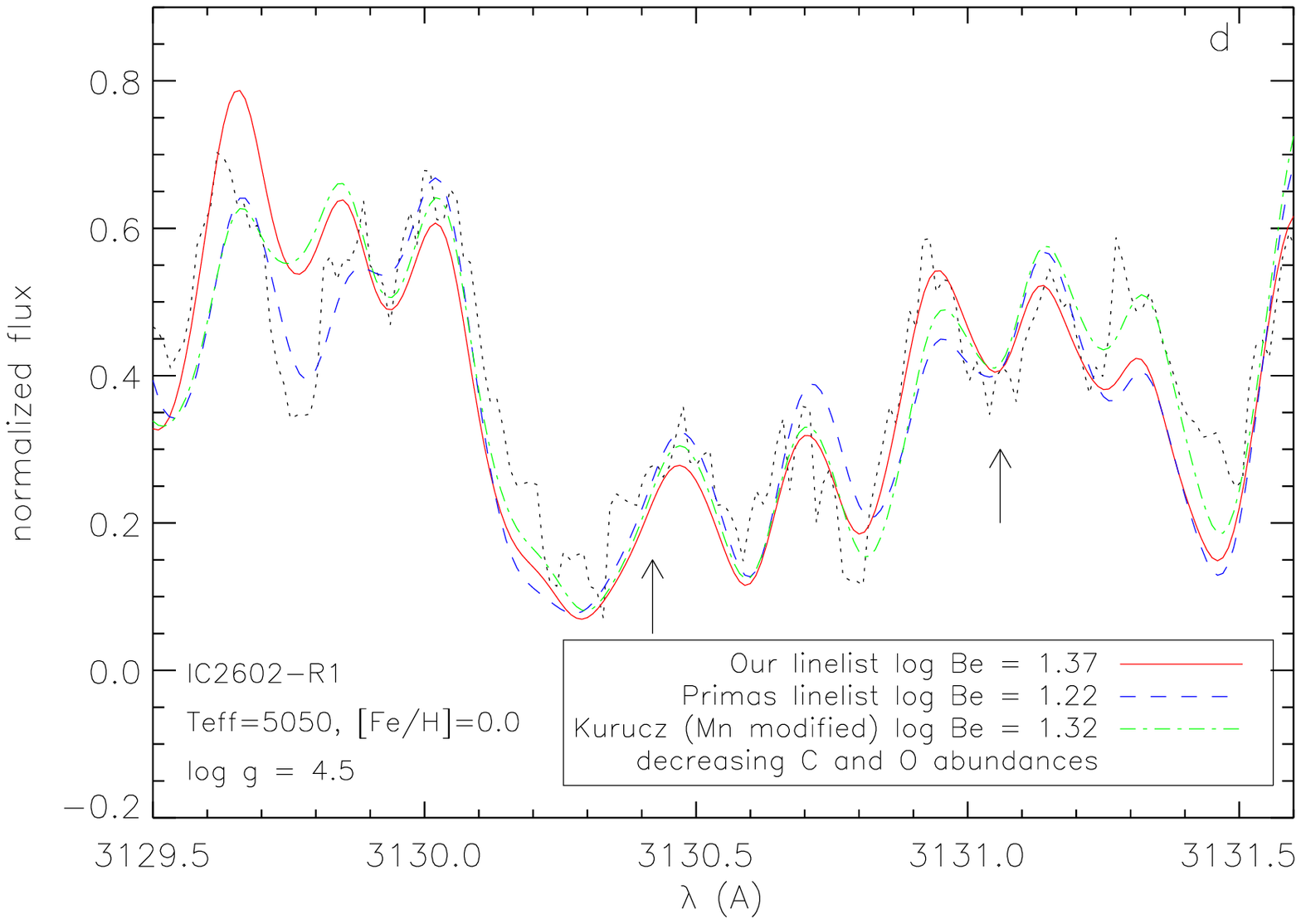}
\caption{Panel a: Spectral synthesis of HD 213042 using different line lists. Panel b: Spectral synthesis of HD 16270 showing 
the contributions of atomic, molecular and Be lines. Panel c: Spectral synthesis of HD 63454 with solar line list and different
values of Be abundance. Panel d: Spectral synthesis of the star R1 in the young cluster IC 2602 using different line lists.}
\label{be_test}
\end{figure*}

\section{Observations and spectral synthesis}

In this study we obtained high resolution spectra for 15 new stars with
magnitudes V between 6 and 10 using the UVES spectrograph at the 8.2-m Kueyen
VLT (UT2) telescope (run ID 86.D-0082A) between October 2010 and March 2011. The
dichroic mirror was used to obtain also red spectra and the slit width was 0.5
arcsec. These new spectra have a spectral resolution \textit{R} $\sim$ 70000 and
\textit{S/N} ratios between 100 and 200. All the data were reduced with the
pipeline of \textit{UVES/VLT}. Standard background correction, flat-field and
extraction procedures were used. The wavelength calibration was made using a
ThAr lamp spectrum taken during the same night. Finally we manually normalized
the continuum by dividing the spectra by a spline function with three
pieces (and the parameter $low-reject$ set to 1) in the whole blue region (3040\AA{}-3800\AA{}).
This normalization does not present any bias for stars with and without planets. When plotting 
together observed spectra of stars with and without planets we only had to multiply the flux of
comparison stars by 0.9-1.1 in order to make them match up. We note that in our previous works we have analyzed high metallicity (up to 0.4) solar type stars with Teff down to 5300 K
 for which we could make a good normalization. The main problem is the normalization of spectra of stars with Teff less
than $\sim$ 5300 K. These stars could have spots and inhomogeneous atmospheres  that make the spectral
synthesis more difficult.\\

The uniform stellar atmospheric parameters were taken from \citet{sousa08} with
typical errors of 25 K for $T_{\rm eff}$, 0.04 dex for $\log g$, 0.03 km s$^{\rm
-1}$ for $\xi_{t}$ and 0.02 dex for metallicity. We refer to that work for
further details in these parameters and their uncertainties.\\

Be abundances were derived by fitting the spectral region around the
\ion{Be}{2} line at 3131.06 \AA{} and then using the \ion{Be}{2} line 3130.42
\AA{} to check the consistency of the fit. We used an empirical line list from
\citet{Garcia-Lopez95} tuned to reproduce the solar spectrum (see Table \ref{lista}).
These synthetic spectra were convolved with a rotational profile. We made a standard LTE
analysis with the revised version of the spectral synthesis code MOOG2002
\citep{sneden} and a grid of Kurucz ATLAS9 atmospheres with overshooting
\citep{kur93}. Examples of synthetic spectra and the parameters used 
in the synthesis are shown in Figure \ref{ajustes_be}.\\


The final sample is composed of 70 and 30 stars with and without planets,
respectively, from \citet{santos_be1,santos_be3,santos_be2,galvez}, 14 stars
with planets from \citet{Delgado11} and 5 and 10 stars with and without detected
planets, respectively, from this work. This gives a total sample of 89 stars
with planets and 40 comparison sample stars. All Be abundances for these
89+40 stars were analyzed by our team using the same methodology,
making this a very uniform sample.\\

\begin{deluxetable}{rrrrrrrr}

\tablecaption{Kurucz line list tuned to reproduce solar spectrum.\label{lista}}
\tablewidth{0pt}
\tablehead{
\colhead{$\lambda (\AA)$} & \colhead{Atomic number} & \colhead{$\chi$ (eV)} &
\colhead{$gf$} & \colhead{$\lambda (\AA)$} & \colhead{Atomic number} & \colhead{$\chi$ (eV)} &
\colhead{$gf$}} 
\startdata

  3127.968 &    108.0 &     1.670 &  0.461E-03 & \vline \hspace{0.35cm} 3130.420  &     4.1  &    0.000 &  0.670E+00  \\
  3128.060 &    108.0 &     0.541 &  0.376E-02 & \vline \hspace{0.35cm} 3130.433  &   108.0  &    1.756 &  0.428E-02  \\
  3128.101 &    108.0 &     0.210 &  0.131E-02 & \vline \hspace{0.35cm} 3130.439  &    25.0  &    3.772 &  0.303E-02  \\
  3128.154 &    108.0 &     1.599 &  0.101E-03 & \vline \hspace{0.35cm} 3130.473  &   108.0  &    1.609 &  0.603E-03  \\
  3128.166 &     25.1 &     6.914 &  0.190E-02 & \vline \hspace{0.35cm} 3130.476  &    26.0  &    3.573 &  0.092E-02  \\
  3128.172 &     25.0 &     7.822 &  0.228E-03 & \vline \hspace{0.35cm} 3130.549  &    25.1  &    6.494 &  0.714E-01  \\
  3128.237 &    108.0 &     0.442 &  0.475E-03 & \vline \hspace{0.35cm} 3130.562  &    26.1  &    3.768 &  0.612E-05  \\
  3128.269 &     21.1 &     7.424 &  0.675E+00 & \vline \hspace{0.35cm} 3130.569  &    24.1  &    5.330 &  0.349E-02  \\
  3128.286 &    108.0 &     0.210 &  0.104E-01 & \vline \hspace{0.35cm} 3130.570  &   108.0  &    0.683 &  0.298E-01  \\
  3128.289 &    108.0 &     0.442 &  0.728E-03 & \vline \hspace{0.35cm} 3130.575  &    23.0  &    1.218 &  0.543E-03  \\
  3128.304 &     23.1 &     2.376 &  0.134E+00 & \vline \hspace{0.35cm} 3130.577  &    73.0  &    1.394 &  0.117E+01  \\
  3128.307 &    607.0 &     0.513 &  0.288E-04 & \vline \hspace{0.35cm} 3130.585  &    24.0  &    3.556 &  0.108E-01  \\
  3128.308 &    607.0 &     0.513 &  0.249E-04 & \vline \hspace{0.35cm} 3130.637  &    25.0  &    4.268 &  0.982E-01  \\
  3128.356 &    108.0 &     1.714 &  0.245E-02 & \vline \hspace{0.35cm} 3130.648  &   106.0  &    0.034 &  0.02009    \\
  3128.377 &    108.0 &     1.714 &  0.333E-03 & \vline \hspace{0.35cm} 3130.780  &    41.1  &    0.439 &  0.257E+01  \\
  3128.393 &     77.0 &     1.728 &  0.110E+00 & \vline \hspace{0.35cm} 3130.791  &    45.0  &    0.431 &  0.776E-02  \\
  3128.394 &    106.0 &     0.558 &  0.731E-01 & \vline \hspace{0.35cm} 3130.803  &    22.1  &    0.012 &  0.589E-01  \\
  3128.394 &    106.0 &     0.558 &  0.800E-01 & \vline \hspace{0.35cm} 3130.813  &    64.1  &    1.157 &  0.826E+00  \\
  3128.406 &     66.1 &     1.314 &  0.226E+01 & \vline \hspace{0.35cm} 3130.842  &    23.0  &    1.955 &  0.119E-02  \\
  3128.488 &     22.1 &     7.867 &  0.151E+01 & \vline \hspace{0.35cm} 3130.851  &    26.2  &   11.595 &  0.527E-03  \\
  3128.495 &     22.1 &     2.590 &  0.908E-05 & \vline \hspace{0.35cm} 3130.871  &    58.1  &    1.090 &  0.957E-01  \\
  3128.518 &   108.0  &    0.786  & 0.817E-02  & \vline \hspace{0.35cm} 3130.905  &    26.1  &    7.487 &  0.100E-02  \\
  3128.524 &   108.0  &    0.102  & 0.500E-03  & \vline \hspace{0.35cm} 3130.928  &   106.0  &    0.002 &  0.231E-03  \\
  3128.546 &    24.1  &    4.757  & 0.859E-02  & \vline \hspace{0.35cm} 3130.928  &   108.0  &    1.907 &  0.946E-03  \\
  3128.568 &    64.1  &    1.134  & 0.787E+00  & \vline \hspace{0.35cm} 3130.933  &   108.0  &    0.683 &  0.294E-03  \\
  3128.617 &    22.0  &    5.959  & 0.658E-03  & \vline \hspace{0.35cm} 3130.997  &   108.0  &    1.569 &  0.137E-03  \\
  3128.618 &    22.0  &    1.067  & 0.883E-03  & \vline \hspace{0.35cm} 3131.015  &    25.1  &    6.112 &  0.607E-01  \\
  3128.626 &    22.0  &    6.065  & 0.643E+00  & \vline \hspace{0.35cm} 3131.037  &    25.0  &    3.773 &  0.596E+00  \\
  3128.641 &    25.1  &    6.672  & 0.714E-01  & \vline \hspace{0.35cm} 3131.059  &    25.1  &    6.672 &  0.191E-02  \\
  3128.648 &    26.1  &   12.966  & 0.855E-02  & \vline \hspace{0.35cm} 3131.065  &     4.1  &    0.000 &  0.338E+00  \\
  3128.653 &   607.0  &    0.510  & 0.249E-04  & \vline \hspace{0.35cm} 3131.070  &    90.1  &    0.000 &  0.276E-01  \\
  3128.653 &   607.0  &    0.510  & 0.210E-04  & \vline \hspace{0.35cm} 3131.102  &    26.1  &    9.688 &  0.695E-01  \\
  3128.692 &    24.1  &    2.434  & 0.479E+00  & \vline \hspace{0.35cm} 3131.109  &    40.0  &    0.520 &  0.398E+00  \\
  3128.692 &    29.0  &    4.974  & 0.195E+00  & \vline \hspace{0.35cm} 3131.115  &    26.0  &    3.047 &  0.194E-05  \\
  3128.694 &    23.1  &    2.372  & 0.430E+00  & \vline \hspace{0.35cm} 3131.116  &    76.0  &    1.841 &  0.112E+01  \\
  3128.728 &    28.0  &    1.951  & 0.512E-04  & \vline \hspace{0.35cm} 3131.143  &    22.0  &    0.836 &  0.279e-05  \\
  3128.737 &    39.1  &    3.376  & 0.646E+01  & \vline \hspace{0.35cm} 3131.194  &    42.0  &    2.499 &  0.441E-01  \\
  3128.763 &    72.0  &    0.000  & 0.170E-01  & \vline \hspace{0.35cm} 3131.212  &    24.0  &    3.113 &  0.604E+00  \\
  3128.776 &    23.0  &    1.804  & 0.126E-01  & \vline \hspace{0.35cm} 3131.243  &    26.0  &    2.176 &  1.726E-04  \\
  3128.782 &   108.0  &    0.897  & 0.102E-01  & \vline \hspace{0.35cm} 3131.255  &    69.1  &    0.000 &  0.240E+00  \\
  3128.854 &    23.0  &    1.712  & 0.205E-03  & \vline \hspace{0.35cm} 3131.326  &    27.1  &    2.204 &  0.817E-04  \\
  3128.898 &    26.0  &    1.557  & 0.223E-02  & \vline \hspace{0.35cm} 3131.329  &   108.0  &    1.942 &  0.160E-01  \\
  3128.938 &   108.0  &    1.939  & 0.254E-03  & \vline \hspace{0.35cm} 3131.338  &    25.0  &    4.679 &  0.117E-01  \\
  3128.949 &    75.0  &    2.061  & 0.166E+01  & \vline \hspace{0.35cm} 3131.339  &    21.1  &    7.381 &  0.372E-02  \\
  3128.954 &    25.0  &    2.920  & 0.447E-04  & \vline \hspace{0.35cm} 3131.366  &   108.0  &    1.942 &  0.250E-03  \\
  3128.975 &   108.0  &    1.939  & 0.169E-01  & \vline \hspace{0.35cm} 3131.384  &   108.0  &    1.680 &  0.604E-03  \\
  3129.001 &   607.0  &    0.508  & 0.210E-04  & \vline \hspace{0.35cm} 3131.394  &   108.0  &    1.680 &  0.511E-03  \\
  3129.005 &    27.0  &    0.514  & 0.117E-02  & \vline \hspace{0.35cm} 3131.395  &    26.1  &    3.815 &  0.221E-03  \\
  3129.009 &    26.1  &    3.968  & 0.202E-02  & \vline \hspace{0.35cm} 3131.423  &   108.0  &    0.960 &  0.789E-02  \\
  3129.009 &    26.1  &   12.966  & 0.340E-02  & \vline \hspace{0.35cm} 3131.458  &    25.1  &    4.340 &  0.294E-04  \\
  3129.013 &    24.1  &   12.978  & 0.873E+00  & \vline \hspace{0.35cm} 3131.459  &    26.0  &    6.427 &  0.234E-04  \\
  3129.017 &   107.0  &    0.740  & 0.202E-04  & \vline \hspace{0.35cm} 3131.486  &    28.1  &   12.409 &  0.755E-02  \\
  3129.038 &    26.2  &   10.311  & 0.229E-03  & \vline \hspace{0.35cm} 3131.502  &   108.0  &    0.494 &  0.185E-02  \\
  3129.070 &    22.0  &    6.079  & 0.971E+00  & \vline \hspace{0.35cm} 3131.525  &    28.0  &    7.152 &  0.126E-02  \\
  3129.095 &   108.0  &    0.897  & 0.574E-03  & \vline \hspace{0.35cm} 3131.533  &    24.1  &    4.168 &  0.782E-02  \\
  3129.110 &    20.0  &    4.625  & 0.185E-02  & \vline \hspace{0.35cm} 3131.545  &    80.0  &    4.887 &  0.912E+00  \\
  3129.138 &   107.0  &    0.740  & 0.211E-04  & \vline \hspace{0.35cm} 3131.548  &    24.1  &    4.178 &  0.350E-01  \\
  3129.144 &    24.1  &    7.332  & 0.498E-01  & \vline \hspace{0.35cm} 3131.583  &    25.1  &    6.495 &  0.124E-01  \\
  3129.153 &    40.1  &    0.527  & 0.479E+00  & \vline \hspace{0.35cm} 3131.656  &   107.0  &    0.787 &  0.140E-03  \\
  3129.182 &    26.0  &    6.411  & 0.349E-04  & \vline \hspace{0.35cm} 3131.687  &   108.0  &    1.736 &  0.193E-01  \\
  3129.183 &    22.0  &    1.046  & 0.160E-03  & \vline \hspace{0.35cm} 3131.702  &    28.0  &    7.264 &  0.247E-01  \\
  3129.209 &   107.0  &    0.740  & 0.223E-04  & \vline \hspace{0.35cm} 3131.711  &   108.0  &    1.736 &  0.319E-03  \\
  3129.210 &    24.0  &    3.556  & 0.490E-01  & \vline \hspace{0.35cm} 3131.724  &    26.1  &    4.081 &  0.486E-02  \\
  3129.228 &    76.0  &    2.191  & 0.537E+00  & \vline \hspace{0.35cm} 3131.754  &   108.0  &    0.960 &  0.550E-03  \\
  3129.300 &    28.0  &    0.275  & 0.625E-02  & \vline \hspace{0.35cm} 3131.812  &    72.0  &    1.306 &  0.263E+01  \\
  3129.305 &    66.0  &    0.000  & 0.135E-01  & \vline \hspace{0.35cm} 3131.825  &    27.0  &    1.740 &  0.161E-01  \\
  3129.333 &    26.0  &    1.485  & 0.114E-01  & \vline \hspace{0.35cm} 3131.838  &    80.0  &    4.887 &  0.912E+00  \\
  3129.348 &    25.0  &    3.379  & 0.490E-04  & \vline \hspace{0.35cm} 3131.935  &   107.0  &    0.787 &  0.148E-03  \\
  3129.348 &    25.0  &    3.379  & 0.195E-03  & \vline \hspace{0.35cm} 3132.053  &    24.1  &    2.483 &  0.120E+01  \\
  3129.376 &    11.1  &   32.944  & 0.102E+01  & \vline \hspace{0.35cm} 3132.062  &    22.0  &    5.975 &  0.530E-03  \\
  3129.389 &    23.0  &    2.115  & 0.979E-03  & \vline \hspace{0.35cm} 3132.063  &    40.0  &    0.543 &  0.105E+01  \\
  3129.454 &     8.1  &   25.640  & 0.290E+00  & \vline \hspace{0.35cm} 3132.109  &    24.1  &    4.775 &  0.643E-04  \\
  3129.478 &    23.1  &    8.574  & 0.152E+00  & \vline \hspace{0.35cm} 3132.142  &    23.0  &    0.262 &  0.589E-05  \\
  3129.481 &    27.0  &    1.883  & 0.951E-02  & \vline \hspace{0.35cm} 3132.186  &   108.0  &    0.901 &  0.964E-02  \\
  3129.538 &   108.0  &    0.516  & 0.178E-02  & \vline \hspace{0.35cm} 3132.193  &   107.0  &    0.787 &  0.160E-03  \\
  3129.548 &    73.0  &    1.147  & 0.107E+00  & \vline \hspace{0.35cm} 3132.212  &    27.0  &    0.101 &  0.377E-01  \\
  3129.589 &    72.0  &    0.000  & 0.135E-01  & \vline \hspace{0.35cm} 3132.281  &   106.0  &    0.488 &  0.661E-01  \\
  3129.636 &    22.0  &    1.443  & 0.158E-02  & \vline \hspace{0.35cm} 3132.281  &   106.0  &    0.488 &  0.592E-01  \\
  3129.652 &    41.1  &    1.321  & 0.115E+00  & \vline \hspace{0.35cm} 3132.288  &    25.0  &    4.332 &  0.316E+00  \\
  3129.763 &    40.1  &    0.039  & 0.331E+00  & \vline \hspace{0.35cm} 3132.355  &    23.0  &    1.043 &  0.984E-04  \\
  3129.774 &    24.0  &    2.708  & 0.344E-02  & \vline \hspace{0.35cm} 3132.392  &   108.0  &    1.990 &  0.131E-03  \\
  3129.857 &    24.0  &    2.968  & 0.845E-02  & \vline \hspace{0.35cm} 3132.405  &    25.0  &    3.373 &  0.818E-02  \\
  3129.934 &    39.1  &    3.414  & 0.955E+01  & \vline \hspace{0.35cm} 3132.517  &    68.1  &    1.402 &  0.320E+01  \\
  3129.937 &   108.0  &    1.609  & 1.955E-02  & \vline \hspace{0.35cm} 3132.518  &    26.0  &    7.168 &  0.316E+00  \\
  3129.943 &    73.0  &    0.697  & 0.724E-01  & \vline \hspace{0.35cm} 3132.532  &    24.1  &    6.805 &  0.387E-03  \\
  3129.968 &    64.1  &    1.172  & 0.627E+00  & \vline \hspace{0.35cm} 3132.579  &    26.1  &    7.495 &  0.161E-02  \\
  3129.974 &    90.1  &    1.287  & 0.172E+00  & \vline \hspace{0.35cm} 3132.583  &   108.0  &    1.612 &  0.314E-01  \\
  3130.056 &    40.0  &    0.519  & 0.200E+00  & \vline \hspace{0.35cm} 3132.591  &    58.1  &    0.295 &  0.288E+00  \\
  3130.063 &    26.1  &   13.018  & 0.416E-02  & \vline \hspace{0.35cm} 3132.594  &    42.0  &    0.000 &  0.237E+01  \\
  3130.075 &   108.0  &    2.295  & 0.0029174  & \vline \hspace{0.35cm} 3132.596  &    23.1  &    2.900 &  0.859E-01  \\
  3130.126 &   108.0  &    0.842  & 0.794E-02  & \vline \hspace{0.35cm} 3132.631  &   607.0  &    0.510 &  0.210E-04  \\
  3130.145 &   108.0  &    1.987  & 0.0141579  & \vline \hspace{0.35cm} 3132.656  &    73.0  &    0.491 &  0.110E+00  \\
  3130.157 &    22.0  &    5.941  & 0.346E+00  & \vline \hspace{0.35cm} 3132.657  &    27.0  &    2.878 &  0.240E-04  \\
  3130.202 &    25.1  &    4.801  & 0.193E+00  & \vline \hspace{0.35cm} 3132.710  &    22.0  &    5.954 &  0.171E+00  \\
  3130.254 &   106.0  &    0.521  & 0.0660693  & \vline \hspace{0.35cm} 3132.725  &    25.1  &    6.177 &  0.215E-02  \\
  3130.257 &    23.1  &    0.348  & 0.513E+00  & \vline \hspace{0.35cm} 3132.788  &    25.0  &    4.268 &  0.316E+00  \\
  3130.281 &   108.0  &    0.250  & 0.134E-01  & \vline \hspace{0.35cm} 3132.794  &     8.2  &   36.895 &  0.933E+00  \\
  3130.290 &   106.0  &    0.521  & 0.0731139  & \vline \hspace{0.35cm} 3132.809  &    23.1  &    2.510 &  0.297E-01  \\
  3130.340 &    58.1  &    0.529  & 0.705E+00  & \vline \hspace{0.35cm} 3132.816  &   108.0  &    1.947 &  0.313E-03  \\
  3130.353 &    27.1  &    2.985  & 0.465E-03  & \vline \hspace{0.35cm} 3132.822  &    24.0  &    3.122 &  0.322E+00  \\
  3130.370 &   106.0  &    0.033  & 0.111E-01  & \vline \hspace{0.35cm} 3132.845  &   108.0  &    1.947 &  0.348E-02  \\
  3130.376 &    22.0  &    1.430  & 0.275E-01  & \vline \hspace{0.35cm} 3132.864  &    28.1  &    2.865 &  0.223E-03  \\
  3130.407 &   108.0  &    1.756  & 0.324E-03  & \vline \hspace{0.35cm} 3132.865  &   108.0  &    0.686 &  0.578E-03  \\
  3130.408 &   108.0  &    1.670  & 0.103E-03  & \vline \hspace{0.35cm} 3132.878  &    44.0  &    1.317 &  0.174E+00  \\
                                                                                              
\enddata
\tablenotetext{*}{Line list used by \citet{Garcia-Lopez95} and in our previous works. The dissociation
energy for OH, CH, NH and CN molecules is 4.39, 3.46, 3.47 and 7.65 eV respectively.}
\end{deluxetable}

\begin{deluxetable}{lcccccclrr}

\tablecaption{Stars analyzed in this work.\label{tabla}}
\tablewidth{0pt}
\tablehead{
\colhead{Star} & \colhead{T$_{\rm eff}$} & \colhead{log \textit{g}} &
\colhead{$\xi_{t}$} & \colhead{[Fe/H]} & \colhead{V} & \colhead{planet} &
\colhead{Spectral type\tablenotemark{a}} & \colhead{log $\epsilon$(Be)} &
\colhead{log $\epsilon$(Li)}\\
\colhead{} & \colhead{[K]} & \colhead{[cm s$^{-2}$]} & \colhead{[km s$^{-1}$]}}
\startdata

HD2638   &   5198  &   4.43 &  0.74 &  0.12  &    9.44 &    yes &   G5       &  0.49     & $<$0.16 \\
HD8326   &   4971  &   4.48 &  0.81 &  0.02  &    8.70 &     no &   K2V      & <$-0.16$  & $<$0.09 \\ 
HD8389A  &   5283  &   4.37 &  1.06 &  0.34  &    7.84 &     no &   K0VCN+2  & 0.16     & $<$0.73 \\ 
HD9796   &   5179  &   4.38 &  0.66 & -0.25  &    8.81 &     no &   K0V      &  0.27     & $<$0.17 \\ 
HD11964A\tablenotemark{b} &   5332  &   3.90 &  0.99 &  0.08  &    6.42 &    yes&   G9VCN+1  &   0.55     & 1.41 \\ 
HD15337  &   5179  &   4.39 &  0.70 &  0.06  &    9.10 &     no &   K1V      &  0.58     & $<$0.42 \\ 
HD16270  &   4786  &   4.39 &  0.84 &  0.06  &    8.37 &     no &   K3.5Vk:  & $<$-0.32  & $<$0.03 \\ 
HD21019\tablenotemark{b}  &   5468  &   3.93 &  1.05 & -0.45  &    6.20 &     no&   G2V      &   0.22     & 1.39 \\ 
HD27894  &   4952  &   4.39 &  0.78 &  0.20  &    9.42 &    yes &   K2V      & $<$-0.38  & $<$0.22 \\ 
HD35854  &   4928  &   4.46 &  0.54 & -0.13  &    7.74 &     no &   K2V      & $<$-0.31  & $<$-0.22 \\ 
HD40105\tablenotemark{b}  &   5137  &   3.85 &  0.97 &  0.06  &    6.52 &     no&   K1IV-V   &  $<$-0.12  & 1.40 \\ 
HD44573  &   5071  &   4.48 &  0.80 & -0.07  &    8.46 &     no &   K2.5Vk:  & 0.75      & $<$-0.01 \\ 
HD63454  &   4840  &   4.30 &  0.81 &  0.06  &    9.37 &    yes &   K3Vk:    & $<$-0.32  & $<$-0.03 \\ 
HD93083  &   5105  &   4.43 &  0.94 &  0.09  &    8.33 &    yes &   K2IV-V   & $<$-0.14  & $<$0.16 \\ 
HD213042 &   4831  &   4.38 &  0.82 &  0.08  &    7.66 &     no &   K5V      & $<$-0.40  & $<$0.06 \\ 
\enddata
\tablenotetext{a}{Values taken from Simbad}
\tablenotetext{b}{Evolved stars}
\end{deluxetable}

\begin{figure*}[ht!]
\centering
\includegraphics[width=16cm]{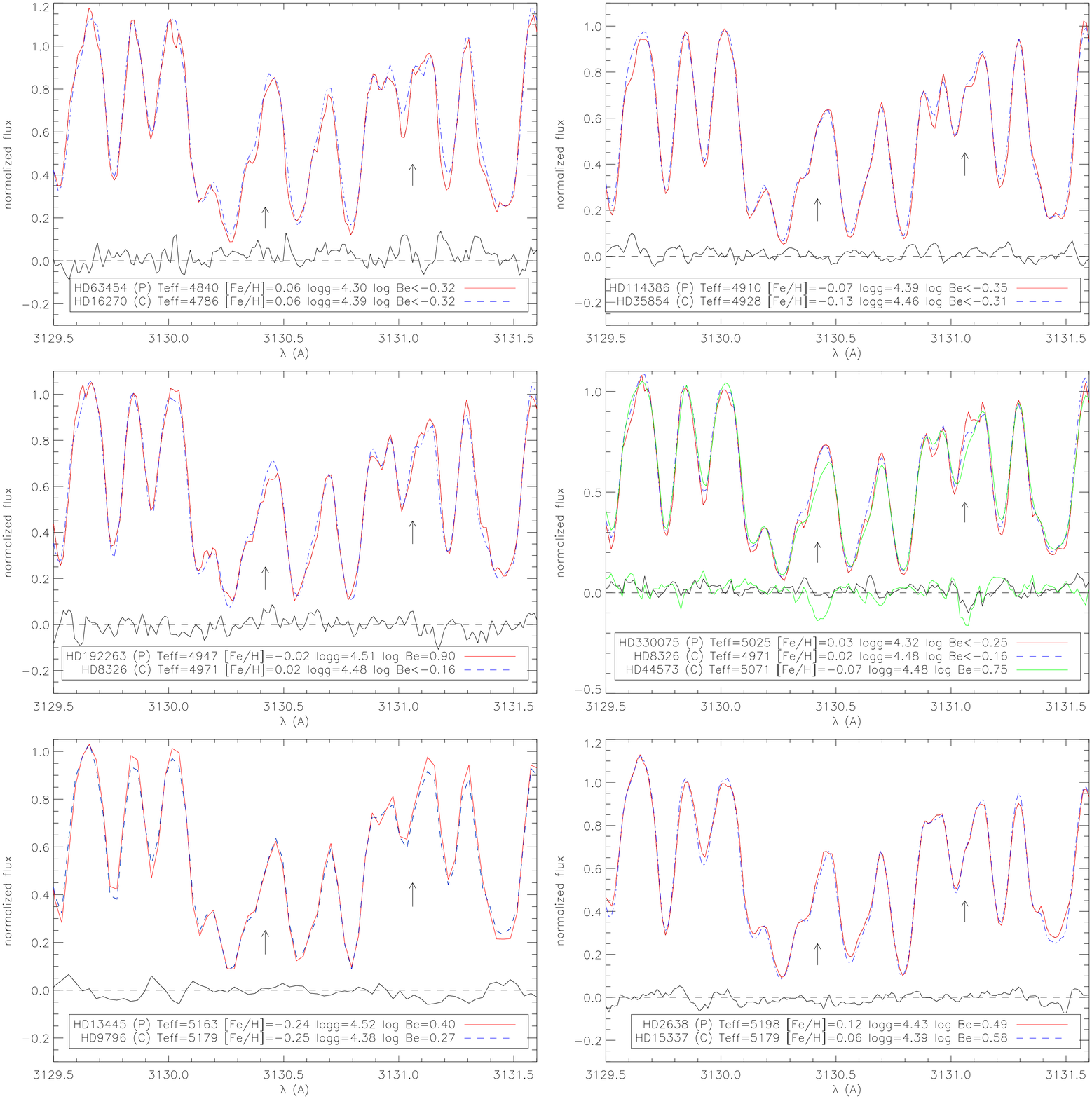}
\caption{Observed spectra and difference in fluxes for six pairs of planet-host
stars (red lines) and stars without detected planets (blue dashed lines). The
position of Be lines are indicated by the arrows.}
\label{be_comp}
\end{figure*}

\section{Analysis}

In general, Be abundances for stars cooler than 5200 K are probably not reliable
since in this regime Be lines are barely sensitive to changes in the abundance. 
In Figure \ref{ajustes_be} we can observe that for the coolest stars the fits are not good.
At those temperatures \ion{Mn}{1} line at 3129.037 \AA{} dominates the feature
and the presence of Be is negligible \citep[see also][]{Garcia-Lopez95}. We make 
several tests to try to improve those fits.

In Figure \ref{be_test} we show several spectral syntheses. In panel \textit{a} we have used
different line lists for HD 213042. The orange dashed line is a fit made with original 
Kurucz line list\footnote{http://kurucz.harvard.edu/line lists.html}. It is clear that 
the Mn-Be feature cannot be well reproduce with the original value of log $gf$ for
Mn line even if we increase Be abundance (light blue dashed-pointed line). Furthermore, 
the lambda of the whole feature do not match Be line, so this star do not present so 
much Be and another line is required to fit the observed spectrum. \citet{primas} used 
an artificial Fe line at 3131.043 which in our case does not help to fit the spectrum 
(purple dashed-three pointed line) since this line gets stronger in metallic stars. 
Another option is to increase the $gf$ of Mn line at 3131.037 \AA{}. This was first proposed 
by \citet{Garcia-Lopez95} to reproduce solar spectrum and we also used this modified line 
in our previous works on Be. 
The green line represents a fit with Kurucz line list and this modified line. We note that  
for our syntheses we have used Mn measured abundances by \citet{neves} with the same models. 
This fit is better but the feature is still stronger than observed in its red wing 
although the Be abundance used in the syntehsis is negligible. 
The synthetic spectra of these cool stars present strong molecular lines. If we decrease 
O and C abundances (red line) we can get a better fit though this does not affect Be line 
at 3131.06 \AA{}.\\

The different contributions of molecular an atomic lines is shown in panel \textit{b} for star
HD 16270. The Mn-Be feature is practically filled with atomic lines, that is, Mn line. 
Therefore we can put an upper limit in Be abundance since increasing Be abundance would 
result in a stronger line with a core shifted towards higher $\lambda$ which would not 
fit the observed spectrum. We can again get a better fit if we decrease O and C abundances 
(red line). However, as the stars get cooler we need to use lower C and O abundances, 
$\sim$ -0.6 dex for O and $\sim$ -0.9 dex for C in this star. This is 
not a realistic approach but fortunately C and O abundances hardly affect the Be feature.\\

In panel \textit{c} we show three syntheses for HD 63454 with the solar line list derived by 
\citet{Garcia-Lopez95} and decreasing C and O abundances -0.3 and -0.2 dex respectively. 
The fits with log Be = 0.6 and 0.3 dex presents a stronger line than observed, 
so we can adopt an upper limit of 0.0 dex for this star. 
Certainly, Be line is not very sensitive at these low temperatures and we think 
that we may be overestimating Be depletion. However, it is imposible to fit the spectra using 
high Be abundances so we can put an upper limit in Be content though we cannot calculate 
accurate abundances. Moreover, the sensitivity of Be line does not seem to be related to 
T$_{\rm eff}$ since some cool young objets present strong Be lines with abundances similar 
to solar \citep{smiljanic,Randich07}. This is the case of the star R1 in the young cluster 
IC 2602 of 46 Myr and solar metallicity. This star was observed by \citet{smiljanic} who 
found log Be = 1.25.
We have analyzed this star using the same spectrum and three different line lists (see 
panel \textit{d} of Figure \ref{be_test}): \citet{primas} line list, used by these authors, gives 
log Be = 1.22; our solar line list gives log Be = 1.37 and Kurucz line list with Mn line 
modified and decreasing C and O abundances by 0.3 dex gives log Be = 1.32, all of them
in perfect agreement with the value previously found. Therefore, this test probably 
suggests that our line list does work for cool stars which present a measurable quantity of Be, 
and our fittings should be valid at least to put an upper limit in Be abundances.\\

\begin{figure*}[ht]

\centering
\includegraphics[width=8cm]{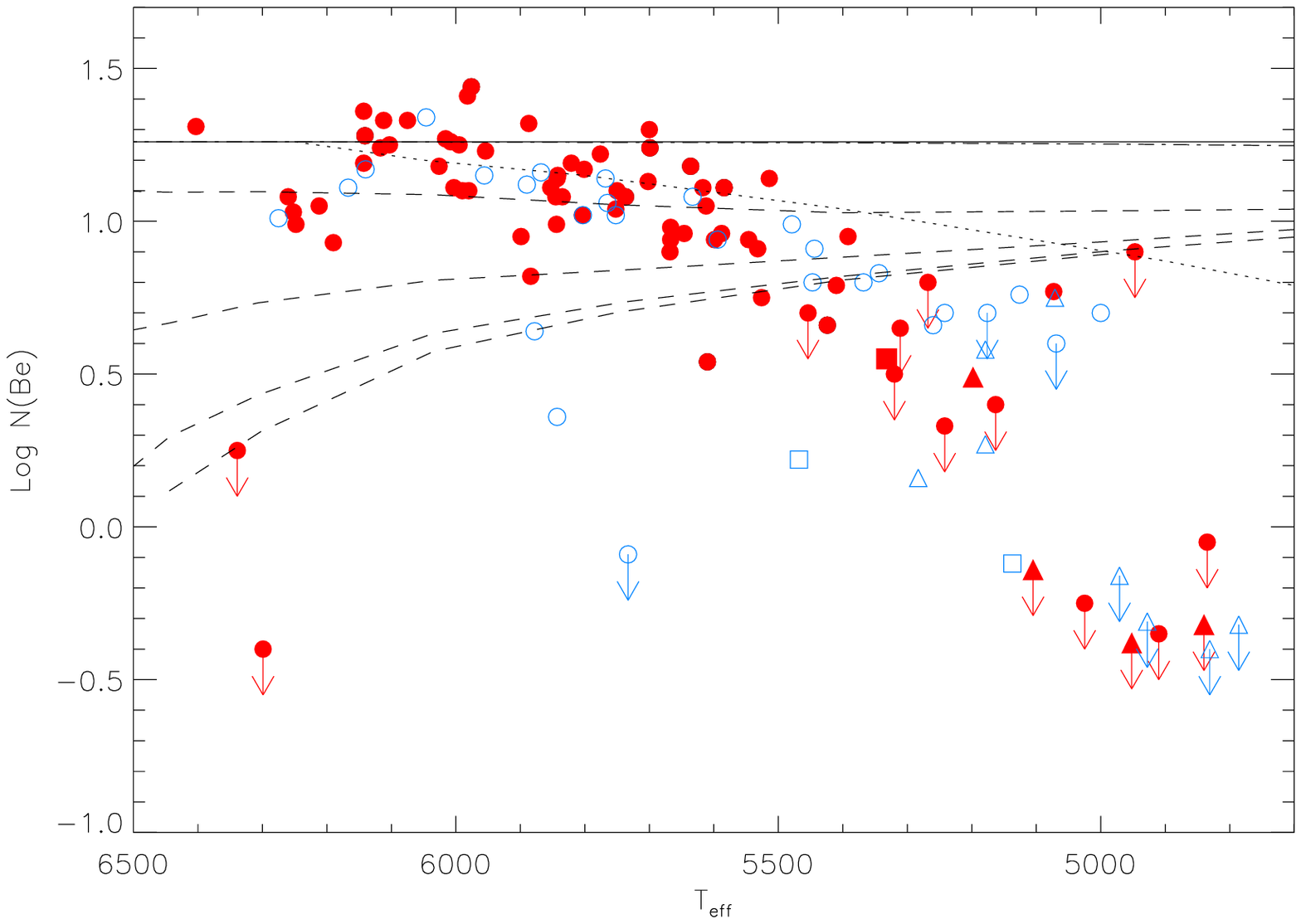}
\includegraphics[width=8cm]{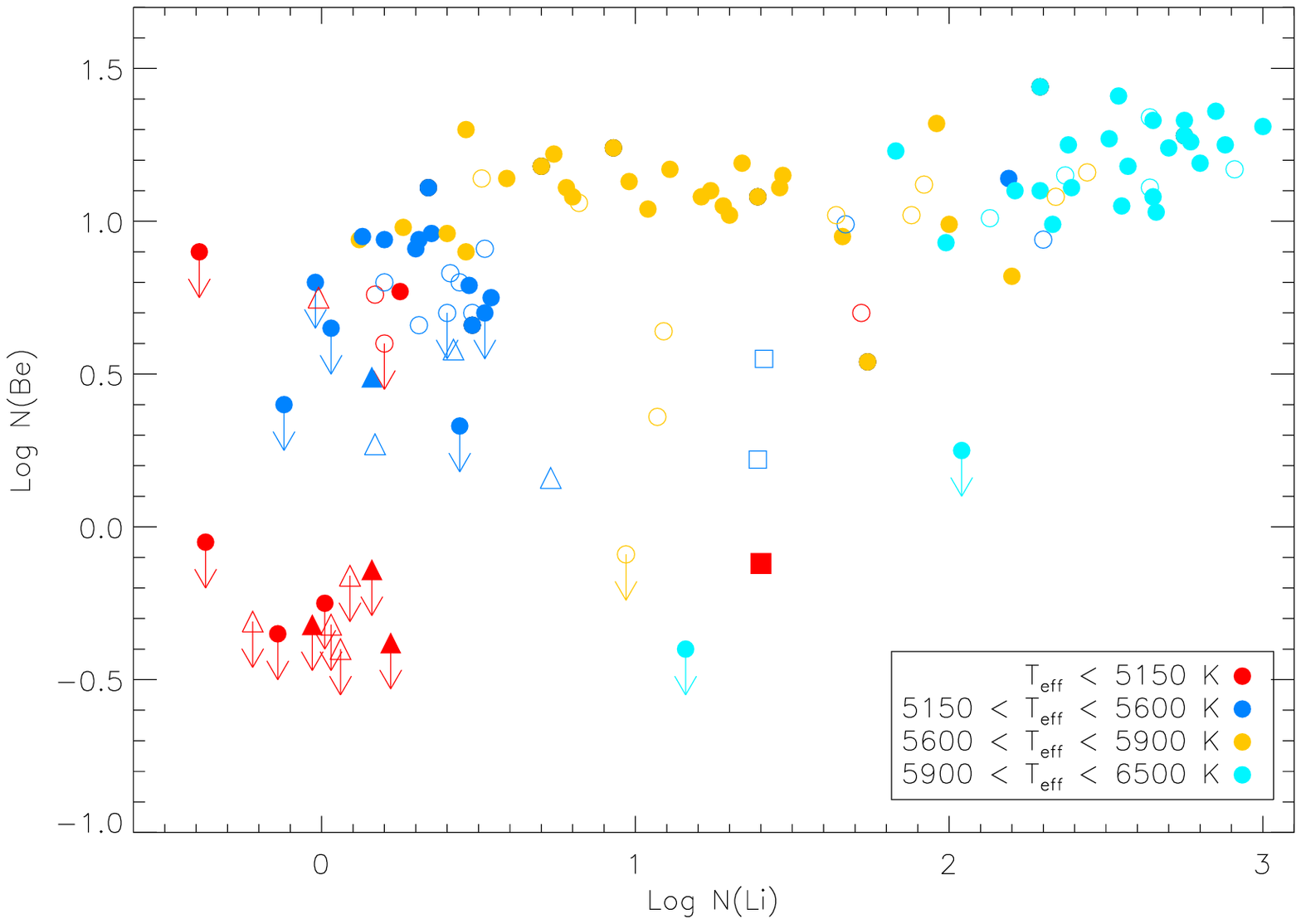}
\caption{\textit{Left panel:} Be abundances as a function of effective
temperature for dwarf stars with and without detected planets from this work
(red filled and blue open triangles, respectively) and dwarf stars with (red
filled circles) and without planets (blue open circles) from previous studies
\citep{santos_be1,santos_be3,santos_be2,galvez,Delgado11}. The three evolved
stars from this work are depicted with squares. The Sun is denoted by
the usual symbol. The dashed lines represent 4 Be depletion models of
\citet{Pinsonneault} (Case A) with different initial angular momentum for solar
metallicity and an age of 1.7 Gyr. The solid line represents an assumed
initial Be abundance of 1.26 \cite{santos_be2}. The dotted line represents the
Be depletion isochrone for 4.6 Gyr taken from the models including mixing by
internal waves of \citet{Montalban}. \textit{Right panel:} Be abundances as a
function of Li abundances. Filled and open circles are stars with 
 and without planets, respectively, from previous surveys. Filled and open
triangles are stars with and without planets, respectively, from this work.
Filled and open squares are evolved stars with and without detected planets,
respectively, with measured Be abundance in this work. Colors denote different
temperature ranges.}
\label{be_teff}
\end{figure*}

\section{Discussion}

In this section we will discuss the different trends of Be abundances
with the effective temperature, the metallicity and the oxygen abundance.

\subsection{Is Be depleted in stars with planets?}

We have seen in the previous section that absolute Be abundances determined with
spectral synthesis are probably not reliable for cool stars, preventing a comparison
between stars with and without detected planets. In order to search for
differences in Be abundances between those stars we proposed to compare
directly their spectra in a previous paper \citep{Delgado11}. We selected pairs
of stars with similar stellar parameters so if there were a difference between
their spectra in the Be region it should be due to a difference in Be abundance.
In that paper we presented two planet-host stars, HD 330075 and HD 13445, with
an extra Be depletion when compared to analogous stars without detected
planets, showing that the effect observed in Li abundances for solar-type stars 
with giant planets might also occur for Be in cooler stars.\\

In this work we present new spectra for 15 cool stars (see Table \ref{tabla})
with and without detected planets. Using these stars and others from previous
works \citep{santos_be3,santos_be2,galvez,Delgado11} we made 13 new pairs of
analogous stars with differences in T$_{\rm eff}$, log \textit{g} and
[Fe/H] lower than 50K, 0.15 dex and 0.10 dex respectively. We also note that v
sin\textit{i} values of these stars are very similar. Some examples of these
couples are shown in Figure \ref{be_comp}. We can see in these plots that stars with
and without planets present similar Be features and the differences in flux
between spectra are very low around Be features. In the previous paper we found
three stars from the comparison sample with clearly higher Be abundance
than the planet-host star HD 330075, whose abundance is log $\epsilon$(Be) $<$
-0.25. Now we have two more stars with very similar parameters. HD 44573 has a
higher Be abundance (see Table \ref{tabla}) and the difference in fluxes between 
both stars is around 5$\sigma$ in the
position of the two Be lines. However, HD 8326 has a spectrum very
similar to HD 330075 and Be abundances are almost equal. Therefore,
planet host stars and comparison stars can have similar Be abundances.
We note, however, that our sample of cool stars is small and although unlikely,
it might be possible that we are only observing the fraction of comparison stars
with a strong Be depletion. A similar situation is seen for solar-type stars
without detected planets, where 50\% have high Li abundances while the other
50\% do have its Li depleted like planet-hosts.\\

\begin{figure*}[ht]
\centering
\includegraphics[width=8cm]{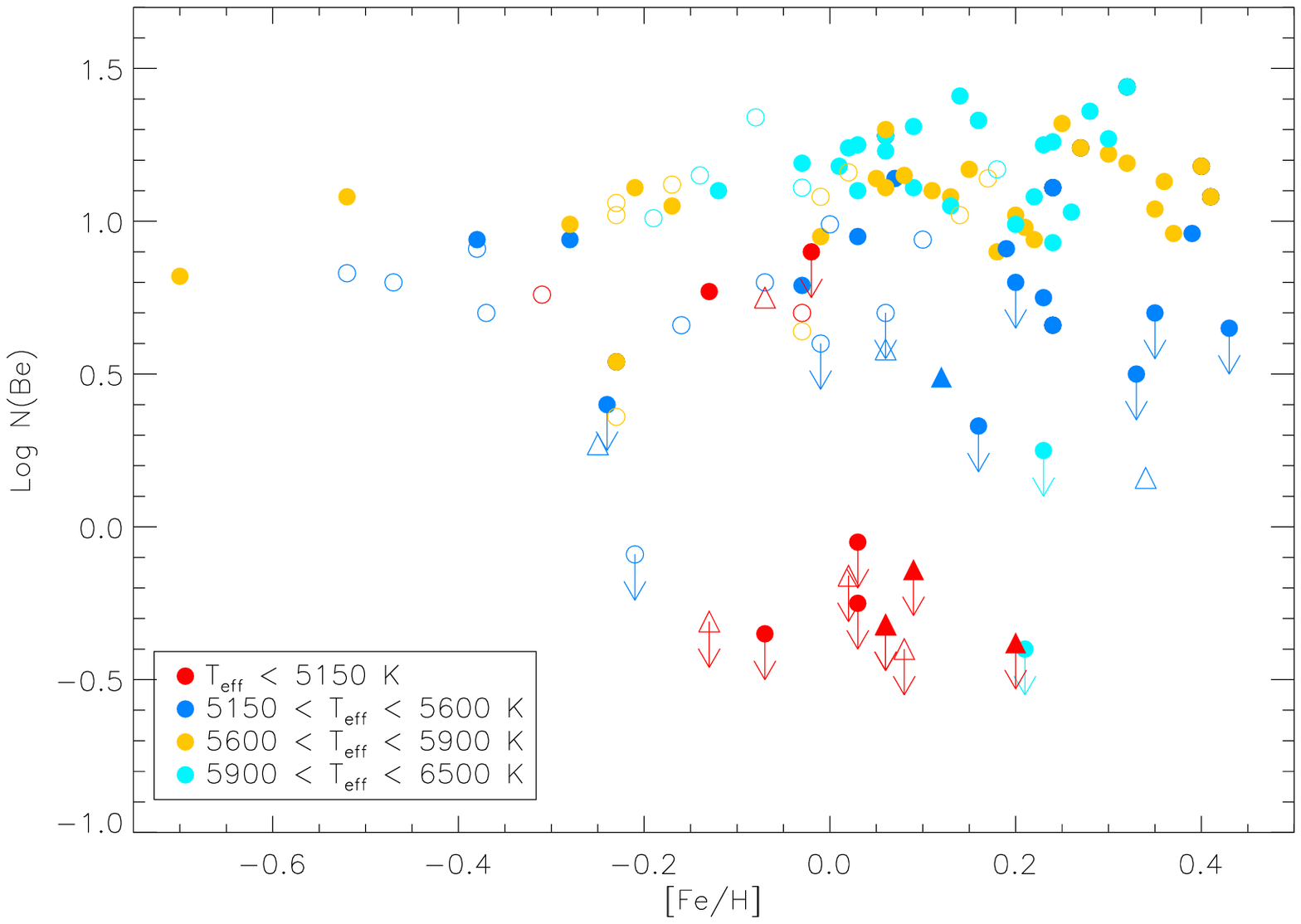}
\includegraphics[width=8cm]{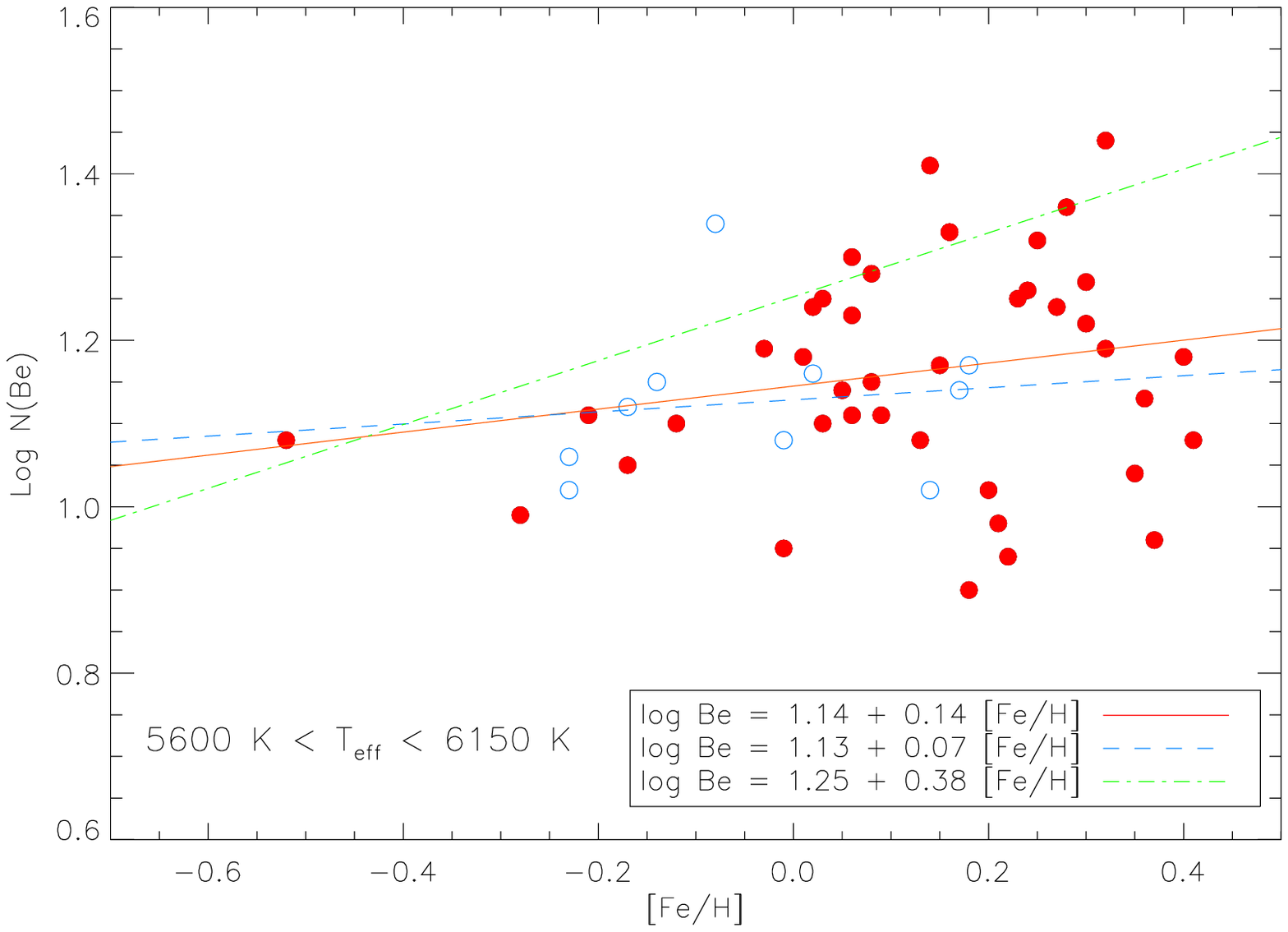}
\caption{\textit{Left panel:} Be abundances as a function of metallicity. Filled
and open circles are dwarf stars with and without planets, respectively, from
previous surveys. Filled and open triangles are dwarf stars with and without
planets, respectively, from this work. Colors denote different temperature
ranges. \textit{Right panel:} Be abundances as a function of metallicity for
dwarf stars with 5600 K $<$ T$_{\rm eff}$ $<$ 6150 K and detections of Be. We
have overplotted two linear least square fittings for planet-host stars (red
filled circles, red line) and stars without detected planets (blue open circles,
blue dashed line).}
\label{be_feh}
\end{figure*}

\begin{figure}[ht]
\centering
\includegraphics[width=8cm]{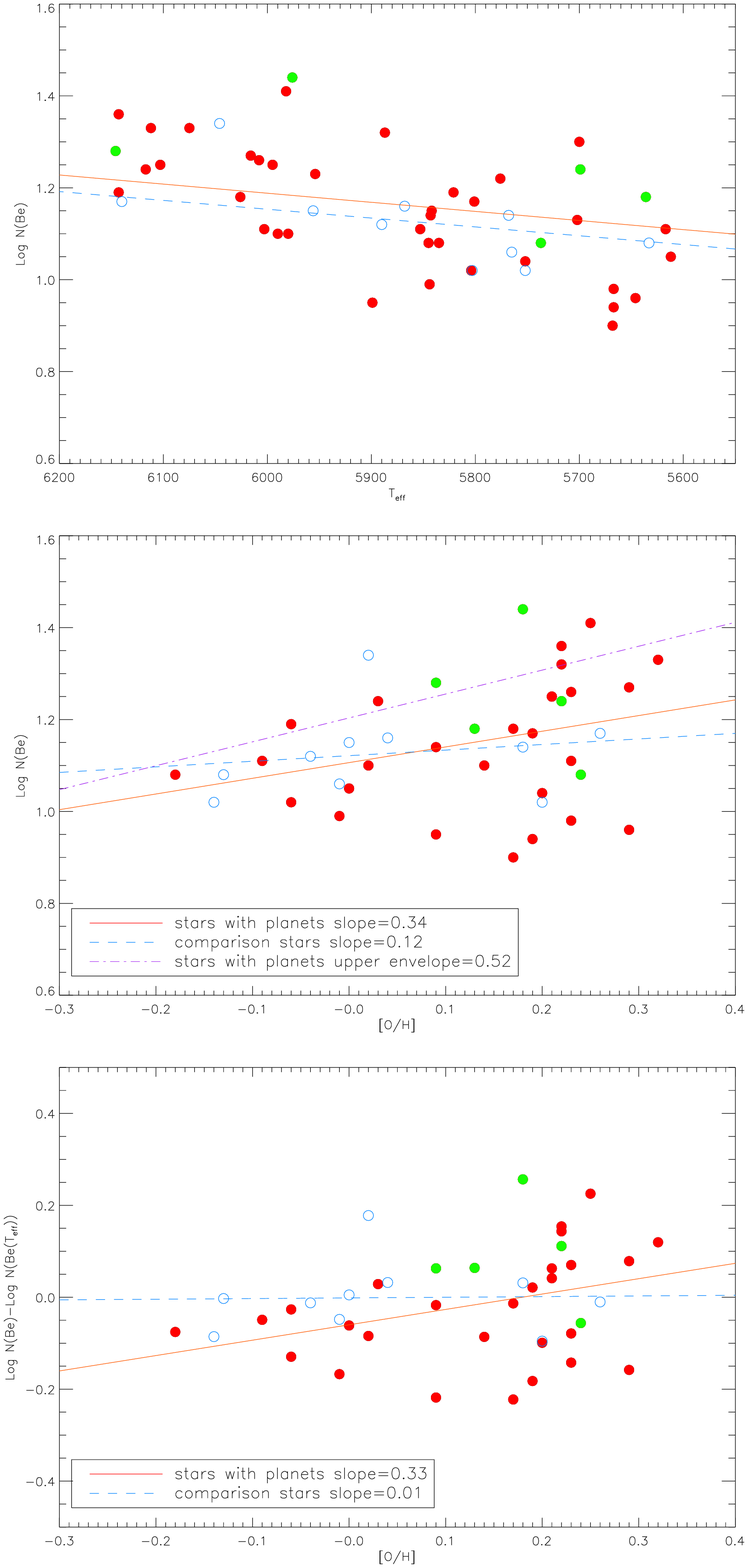}
\caption{\textit{Upper panel:} Be abundances as a function of temperature for
dwarf stars with effective temperature between 5600 K and 6150 K with (red
filled circles) and without planets (blue open circles). Green circles are stars
with planets with oxygen abundances measured in this work. \textit{Middle
panel:} Be abundances as a function of [O/H] using O abundances from OH
\citep{Ecuvillon06} lines for stars with effective temperature between 5400 K
and 6300K. Symbols like in top panel. \textit{Lower panel:} Be abundances
corrected of temperature effect for the same stars than middle panel.}
\label{be_oxi}
\end{figure}

\subsection{Be versus T$_{\rm eff}$}

In left panel of Figure \ref{be_teff} we plot the derived Be abundances as a
function of effective temperature for planet-host stars in our sample (see Table
\ref{tabla}) together with previous samples. In this plot we have removed
subgiants and giants (except the three analyzed in this work) to avoid
evolutionary effects in the abundances. In this selection we took the spectral
types of the stars from \citet{galvez}. With this new sample of stars we have
completed the coolest part of the plot and now we can see the behaviour of Be
abundances in a wide range of effective temperatures.

As mentioned in previous papers, the Be abundances decrease from a maximum near
T$_{\rm eff}$ = 6100 K towards higher and lower temperatures, in a similar way
as Li abundances behave. In the high temperature domain, the steep decrease with
increasing temperatures resembles the well known Be gap for F stars
\citep[e.g.][]{Boesgaard02}. The decrease of the Be content towards lower
temperatures is smoother and may show evidence for continuous Be burning during
the main sequence evolution of these stars.\\ 

When we move to the coolest stars, we observe a strong Be depletion regardless 
of the presence of planets. Hence, the planetary formation processes that 
affect Li depletion in solar analog stars
\citep{israelian09,gonzalez10} do not seem to take effect in the Be abundances.

On the other hand, we have found several stars with undetectable or very weak Be
lines. The steep decrease of Be abundances for T$_{\rm eff}$ $<$ 5500 K is in
contradiction with models of Be depletion \citep{Pinsonneault}, which predict
either constant or increasing Be abundances as T$_{\rm eff}$ decreases (see
Figure \ref{be_teff}). Even taking into account mixing by internal waves
\citep{Montalban}, Be depletion is still lower than predicted. 
Be abundances have also been studied in clusters of different ages. 
\citet{Randich07} found that stars in the 150 Myr old cluster NGC 2516 with 
5000 $<$ T$_{\rm eff}$ $<$ 5500 K have not depleted Be while Hyades stars 
(600 Myr) of the same T$_{\rm eff}$ present some Be depletion, something that
cannot be explained by models with only convective mixing.
\citet{Garcia-Lopez95} also found evidence of Be depletion for 
two cool Hyades members.
Although uncertainties in Be abundandes for the coolest stars are large and it might be
some kind of systematic effect due to the \ion{Mn}{1} $\lambda$~3131.037~{\AA}
line, many of them seem to have their Be totally destroyed. Therefore, none of
those models including convective mixing, rotation or gravitational waves 
seems to fit the observed Be abundances, at least at these cool temperatures. 
Other depletion mechanisms have been proposed to explain Li destruction, 
such as tachocline diffusion \citep{Brun,Piau} or episodic accretion of planetary
material during the PMS \citep{Baraffe} though Be has not been analyzed. However, 
accretion of planetary material in the ZAMS \citep{theado,theado11} can lead to 
an enhanced depletion of Li and Be, as well. These models can explain the 
difference in the Li abundances between stars with and without planets and might 
also explain the observed Be differences in a pair of cool planet-hosts when 
compared to several stars without detected planets \citep{Delgado11}. On the other 
hand the depletion produced by episodic accretion does not seem to be enough to explain the 
strong Be depletion observed in cool stars and the presence of planetary material would be 
necessary in all stars though many of them do not have detected planets.\\

\subsection{Be versus Li}

A beryllium versus lithium diagram can give us information about the depletion
rates in main-sequence stars. In right panel of Figure \ref{be_teff} this
relation is shown for all the stars in our sample. In general Be abundances
increase with increasing Li abundances and stars with and without planets
(filled and open symbols, respectively) behave in a similar way. If we take into
account the effective temperatures we can divide the sample in four groups. The
first one, with the hottest stars (light blue points), present both high Be and
Li abundances except for two stars with T$_{\rm eff}$ around 6300K which fall in
the Be-Li dip region, where the depletion of those elements is atributted to
slow mixing and depends on age and temperature \citep{Boesgaard02,Boesgaard04a},
though one of them, HD 120136 presents a normal Li abundance (log $\epsilon$(Li)
= 2.04). In the second group (orange points), with effective temperatures
between 5600 and 5900 K, Be abundances remain high although Li abundances 
present different depletion rates. There are several
objects with low Be abundances and solar temperatures (specially the comparison
star HD 20766 T$_{\rm eff}$ = 5733 K, log $\epsilon$(Be) $<$ -0.09), which could
form some kind of ``Be-gap'' where the depletion of Be may be related to
different pre-main sequence rotational histories \citep{santos_be2}. In a recent
work, \citet{takeda2011} found four stars with solar temperatures and Be line
undetectable. Those stars are also depleted in Li and have low \textit{v}
sin\textit{i} values indicating that slow rotation can be the cause for that
strong Be depletion.

In the third group (blue points), composed by stars with 5150K $<$ T$_{\rm eff}$
$<$ 5600K, Be abundances begin to decrease and Li is severely depleted except
for two objects, HD 11964A and HD 21019, with particularly high Li abundances
considering its temperature. These two objects are evolved stars, therefore a
possible explanation for the high Li content is that they have just left the
main sequence, where they were hotter than they are now, but there has not been
time for their Li and Be to become strongly depleted. Other possible
explanations are a dredge up effect from a ``buffer'' below the former main
sequence convective envelope \citep{Deliyannis90} or accretion of planetary
metal-rich material, although recent models indicate that accretion of planetary
material might destroy Li instead of producing an enhancement
\citep{Baraffe,theado}. Finally, the coolest objects of our sample (red symbols)
are depleted in both Be and Li, although some stars have preserved some Be. 
There are two objects with anomalous high Li content. HD 74576 seems to be a
young star, something that probably justifies its high Li content \citep[see
discussion in][]{santos_be2} while HD 40105 is an evolved star that might have
suffered some of the processes proposed for HD 11964A and HD 21019. We note that
we have removed from the plots all the evolved stars from previous works. Only
the three subgiants observed for this work are analyzed here but there are
similar cases in the whole sample \citep[see][]{santos_be2,galvez}.\\

This figure confirms what found in \citet{santos_be3,santos_be2,galvez}, Be and
Li burning seems to follow the same trend. Both elements increase their content
as the temperature rises, although Li depletion begins at 5900 K while strong Be
depletion starts below 5500 K. Stars with and without planets behave in a
similar way.\\

















\subsection{Be versus [Fe/H]}

In the left panel of Figure \ref{be_feh} we show Be abundances as a function of
metallicity for different temperature ranges. We can see that the stars are
equally distributed regardless of their temperature and for higher metallicities
the dispersion in Be abundances increases, mainly due to objects with 
5150 K $<$ T$_{\rm eff}$ $<$ 5600 K. It is well known that Be abundances
increase with metallicity for [Fe/H] $<$ -1 with a steep slope near 1
\citep[e.g.][]{Rebolo88,Boesgaard09}, but for higher metallicities this relation
is not so well defined.

\citet{Boesgaard04b} found a slope of 0.38 for stars with metallicity between
-0.6 and 0.2, in agreement with \citet{takeda2011} whose slope value is 0.49,
although they take into account only solar analogs with -0.3 $<$ [Fe/H] $<$ 0.3.
On the other hand, \citet{Boesgaard09} argued that stars of one solar mass with
solar metallicities (their most metallic star has [Fe/H] = 0.11) match the slope
of 0.86 for metal-poor stars.

In the right panel of Figure \ref{be_feh} all the dwarf stars of our sample with
5600 K $<$ T$_{\rm eff}$ $<$ 6150 K are plotted. We have chosen this limit in
temperature in order to remove most of the upper limits in Be abundances and
those stars from the Li-Be dip. We have also removed stars from the solar
Be-gap. We find a slope of 0.13 for planet-host stars and 0.07 for comparison
sample stars. These values are lower than previously found by other authors but
this could be possibly due to the lack of stars with [Fe/H] $<$ -0.2 in our
sample in comparison with the high number of metal-rich objects. It seems that
stars with solar and supersolar metallicities do not follow the same trend as
metal-poor stars, and Be has been produced at a lower rate as the Galaxy has
evolved. However, we note that Be abundances present higher uncertainties in
metal-rich stars that might be affecting these trends. If we have a look in the
upper envelope, which might be more consistent with low
 er metallicity stars (around [Fe/H] = -0.5), we obtain a fit with a slope of
0.38, more consistent with previous results.

\begin{deluxetable}{lrrrrrr}

\tablecaption{Oxygen abundances for stars with T$_{\rm eff}$ $>$ 5400 K from
\citet{Delgado11}.\label{tabla_oxigeno}}

\tablewidth{0pt}

\tablehead{

\colhead{Star} & \colhead{[O/H]$_{1}$} & \colhead{[O/H]$_{2}$} &
\colhead{[O/H]$_{3}$} & \colhead{[O/H]$_{4}$} & \colhead{[O/H]$_{final}$}  &
\colhead{log $\epsilon$(Be)}} 
\startdata
HD2039   &   0.17  &  0.12  &  0.17  &  0.29  &  0.18    & 1.44 \\
HD4203   &   0.08  &  0.01  &  0.22  &  0.20  &  0.13    & 1.18 \\ 
HD73526  &   -     &  -     &  0.20  &  0.25  &  0.22    & 1.24 \\ 
HD76700  &   0.16  &  0.16  &  0.28  &  0.35  &  0.24    & 1.08 \\ 
HD154857 &  -0.23  & -0.20  & -0.18  & -0.18  & -0.20    & 0.54 \\ 
HD208847 &   0.13  &  0.08  &  0.05  &  0.10  &  0.09    & 1.28 \\ 
HD216770 &  -0.01  &  0.04  &  0.06  &  0.06  &  0.04    & 0.66 \\ 
\enddata



\end{deluxetable}

\subsection{Be versus [O/H]}

Beryllium is produced by spallation reactions between galactic cosmic rays
(GCRs) and the CNO nuclei in the interstellar medium \citep[see e.g.][and
references therein]{Tan}. Therefore, oxygen abundances can provide us with
complementary information about the galactic evolution of Be. We compile oxygen
abundances from \citet{Ecuvillon06} \citep[see the final lists in][]{galvez},
derived with the OH bands in the near-UV since this region is closed to Be
lines. We use the same temperature ranges than in previous section, 5600 K $<$
T$_{\rm eff}$ $<$ 6150 K. Thus, none of the stars analyzed in this work are
included in this plot because all of them are cooler than 5600 K. However, 6
planet-host stars from our previous work \citep{Delgado11} are hotter and we
have displayed them as green circles in Figure \ref{be_oxi}. Oxygen abundances
for these stars have been obtained in the same way as \citet{Ecuvillon06} and
are presented in Table \ref{tabla_oxigeno}.\\

In the middle panel of Figure \ref{be_oxi}, Be abundances are plotted as a
function of [O/H]. We can see that the slopes of planet hosts and stars without
detected planets are slightly different, possibly due to the low number of
comparison sample stars in this range. We have removed the star HD154857 since
its O abundance is the lowest of this sample and its content in Be is
considerable lower than other stars of similar oxygen abundance, hence it would
increase the slope by 0.3. We make again a fit of the upper envelope for stars
with planets to take into account the lack of stars with low [O/H] in our plot.
These trends might be influenced by the effect of T$_{\rm eff}$ in Be
abundances. To remove this effect, we have made a new Be vs. [O/H] plot (lower
panel of Figure \ref{be_oxi}) with ``corrected'' Be abundances, calculated as
log $\epsilon$(Be)-log $\epsilon$(Be(T$_{\rm eff}$)), where log
$\epsilon$(Be(T$_{\rm eff}$)) is obtained from a linear fit of Be as a function
o
 f temperature in the same T$_{\rm eff}$ range (upper panel of Figure
\ref{be_oxi}). However, the slope is still the same for stars with planets. it
seems that there remains a correlation between Be and O abundances, which may
indicate that the sensitivity of Be to the Teff is not affecting at all this
correlation. On the other hand, the comparison sample is so small at this
temperature range, that no strong

conclusion can be extracted from the different slopes obtained before and after
the Be-Teff linear correction.

Nevertheless, these slopes are considerable lower than previously found for
metal-poor stars, even the slope of the upper envelope in the middle panel.
\citet{Boesgaard99} obtained a slope of 1.45, using stars with -3 $<$ [Fe/H] $<$
0.05, while \citet{Tan} reported a slope of 1.49, for stars with -2.3 $<$ [Fe/H]
$<$ -0.6. Therefore, the slope for metal-rich stars is much flatter than for
metal-poor stars as it seems to happen for Be versus [Fe/H]. A slope like this
would not be in agreement with models of Be production that predict quadratic or
linear relations \citep[see e.g.][and references therein]{Tan}. We note again
that oxygen abundances present a high dispersion in metal-rich stars and even at
low metallicities, the use of different O indicators change this trend.

\section{Conclusions}

We present new high-resolution UVES/VLT near-UV spectra of 15 stars with and
without planets in order to find possible differences in Be abundances between
them and confirm our previous results that suggested a greater depletion of Be
in planet-host stars when compared with stars without detected planets. Be needs
higher temperatures than Li to be destroyed so we have to search for these
differences in cooler stars, which have deeper convective envelopes and are able
to carry the material towards Be-burning layers. We have made new couples of 
analog stars with and without planets with the purpose
of comparing directly their spectra to see possible differences in Be
abundances. Although in a previous work we found a pair of planet-host stars
with their Be severely depleted, now we have not observed important differences 
in Be abundances between both groups of stars. Thus, the
effect caused by protoplanetary disks and rotational history on extra Li
depletion in solar-type stars with planets is not taking effect in Be
abundances, apparently.\\

Furthermore, Be abundances in cool stars (T$_{\rm eff}$ $<$ 5200 K) are much
lower than predicted by the models. We have analyzed for the first time a
considerable number of cool objects and found a strong destruction of Be in most
of the stars. Although we cannot provide absolute abundances we can give 
upper values for Be. This gives a steep drop of Be abundances as T$_{\rm eff}$
diminishes in contradiction with current models of Be depletion.\\

Finally, the slopes of Be abundances as a function of [Fe/H] and [O/H] are
considerably lower than found for metal-poor stars, indicating that the
production rate of Be may have diminished with the evolution of the Galaxy.

\acknowledgments

E.D.M, J.I.G.H. and G.I. would like to thank financial

support from the Spanish Ministry project MICINN AYA2008-04874.

J.I.G.H. acknowledges financial support from the Spanish Ministry of

Science and Innovation (MICINN) under the 2009 Juan de la Cierva

Programme.\\ 

N.C.S. would like to thank the support by the 

European Research Council/European Community under the FP7 through a 

Starting Grant, as well from Funda\c{c}\~ao para a Ci\^encia e a

Tecnologia (FCT), Portugal, through a Ci\^encia\,2007 

contract funded by FCT/MCTES (Portugal) and POPH/FSE (EC), 

and in the form of grant reference PTDC/CTE-AST/098528/2008

from FCT/MCTES.\\

This research has made use of the SIMBAD database

operated at CDS, Strasbourg, France. \\

This work has also made use of

the IRAF facility, and the Encyclopaedia of extrasolar planets.

\vspace{10cm}

\clearpage























\end{document}